\documentclass[twocolumn]{aastex631}
\usepackage{graphicx}
\usepackage{color}
\usepackage{amsmath}
\usepackage{amssymb}
\usepackage[utf8]{inputenc}

\begin{document}

\title{An in-depth study of Gamma rays from the Starburst Galaxy M\,82 with VERITAS}
\author[0000-0002-2028-9230]{A.~Acharyya}\affiliation{CP3-Origins, University of Southern Denmark, Campusvej 55, 5230 Odense M, Denmark}
\author[0000-0002-9021-6192]{C.~B.~Adams}\affiliation{Physics Department, Columbia University, New York, NY 10027, USA}
\author[0000-0002-3886-3739]{P.~Bangale}\affiliation{Department of Physics and Astronomy and the Bartol Research Institute, University of Delaware, Newark, DE 19716, USA}
\author[0000-0002-9675-7328]{J.~T.~Bartkoske}\affiliation{Department of Physics and Astronomy, University of Utah, Salt Lake City, UT 84112, USA}
\author[0000-0003-2098-170X]{W.~Benbow}\affiliation{Center for Astrophysics $|$ Harvard \& Smithsonian, Cambridge, MA 02138, USA}
\author{Y.~Chen}\affiliation{Department of Physics and Astronomy, University of California, Los Angeles, CA 90095, USA}
\author{J.~L.~Christiansen}\affiliation{Physics Department, California Polytechnic State University, San Luis Obispo, CA 94307, USA}
\author{A.~J.~Chromey}\affiliation{Center for Astrophysics $|$ Harvard \& Smithsonian, Cambridge, MA 02138, USA}
\author[0000-0003-1716-4119]{A.~Duerr}\affiliation{Department of Physics and Astronomy, University of Utah, Salt Lake City, UT 84112, USA}
\author[0000-0002-1853-863X]{M.~Errando}\affiliation{Department of Physics, Washington University, St. Louis, MO 63130, USA}
\author{M.~Escobar~Godoy}\affiliation{Santa Cruz Institute for Particle Physics and Department of Physics, University of California, Santa Cruz, CA 95064, USA}
\author[0000-0002-5068-7344]{A.~Falcone}\affiliation{Department of Astronomy and Astrophysics, 525 Davey Lab, Pennsylvania State University, University Park, PA 16802, USA}
\author{S.~Feldman}\affiliation{Department of Physics and Astronomy, University of California, Los Angeles, CA 90095, USA}
\author[0000-0001-6674-4238]{Q.~Feng}\affiliation{Department of Physics and Astronomy, University of Utah, Salt Lake City, UT 84112, USA}
\author[0000-0002-2944-6060]{J.~Foote}\affiliation{Department of Physics and Astronomy and the Bartol Research Institute, University of Delaware, Newark, DE 19716, USA}
\author[0000-0002-1067-8558]{L.~Fortson}\affiliation{School of Physics and Astronomy, University of Minnesota, Minneapolis, MN 55455, USA}
\author[0000-0003-1614-1273]{A.~Furniss}\affiliation{Santa Cruz Institute for Particle Physics and Department of Physics, University of California, Santa Cruz, CA 95064, USA}
\author[0000-0002-0109-4737]{W.~Hanlon}\affiliation{Center for Astrophysics $|$ Harvard \& Smithsonian, Cambridge, MA 02138, USA}
\author[0000-0002-8513-5603]{D.~Hanna}\affiliation{Physics Department, McGill University, Montreal, QC H3A 2T8, Canada}
\author[0000-0003-3878-1677]{O.~Hervet}\affiliation{Santa Cruz Institute for Particle Physics and Department of Physics, University of California, Santa Cruz, CA 95064, USA}
\author[0000-0001-6951-2299]{C.~E.~Hinrichs}\affiliation{Center for Astrophysics $|$ Harvard \& Smithsonian, Cambridge, MA 02138, USA and Department of Physics and Astronomy, Dartmouth College, 6127 Wilder Laboratory, Hanover, NH 03755 USA}
\author[0000-0002-6833-0474]{J.~Holder}\affiliation{Department of Physics and Astronomy and the Bartol Research Institute, University of Delaware, Newark, DE 19716, USA}
\author[0000-0002-1432-7771]{T.~B.~Humensky}\affiliation{Department of Physics, University of Maryland, College Park, MD, USA and NASA GSFC, Greenbelt, MD 20771, USA}
\author[0000-0002-1089-1754]{W.~Jin}\affiliation{Department of Physics and Astronomy, University of California, Los Angeles, CA 90095, USA}
\author[0009-0008-2688-0815]{M.~N.~Johnson}\affiliation{Santa Cruz Institute for Particle Physics and Department of Physics, University of California, Santa Cruz, CA 95064, USA}
\author[0000-0002-3638-0637]{P.~Kaaret}\affiliation{Department of Physics and Astronomy, University of Iowa, Van Allen Hall, Iowa City, IA 52242, USA}
\author{M.~Kertzman}\affiliation{Department of Physics and Astronomy, DePauw University, Greencastle, IN 46135-0037, USA}
\author{M.~Kherlakian}\affiliation{DESY, Platanenallee 6, 15738 Zeuthen, Germany}
\author[0000-0003-4785-0101]{D.~Kieda}\affiliation{Department of Physics and Astronomy, University of Utah, Salt Lake City, UT 84112, USA}
\author[0000-0002-4260-9186]{T.~K.~Kleiner}\affiliation{DESY, Platanenallee 6, 15738 Zeuthen, Germany}
\author[0000-0002-4289-7106]{N.~Korzoun}\affiliation{Department of Physics and Astronomy and the Bartol Research Institute, University of Delaware, Newark, DE 19716, USA}
\author{F.~Krennrich}\affiliation{Department of Physics and Astronomy, Iowa State University, Ames, IA 50011, USA}
\author[0000-0002-5167-1221]{S.~Kumar}\affiliation{Department of Physics, University of Maryland, College Park, MD, USA }
\author[0000-0003-4641-4201]{M.~J.~Lang}\affiliation{School of Natural Sciences, University of Galway, University Road, Galway, H91 TK33, Ireland}
\author[0000-0003-3802-1619]{M.~Lundy}\affiliation{Physics Department, McGill University, Montreal, QC H3A 2T8, Canada}
\author[0000-0001-9868-4700]{G.~Maier}\affiliation{DESY, Platanenallee 6, 15738 Zeuthen, Germany}
\author[0000-0001-7106-8502]{M.~J.~Millard}\affiliation{Department of Physics and Astronomy, University of Iowa, Van Allen Hall, Iowa City, IA 52242, USA}
\author[0000-0001-5937-446X]{C.~L.~Mooney}\affiliation{Department of Physics and Astronomy and the Bartol Research Institute, University of Delaware, Newark, DE 19716, USA}
\author[0000-0002-1499-2667]{P.~Moriarty}\affiliation{School of Natural Sciences, University of Galway, University Road, Galway, H91 TK33, Ireland}
\author[0000-0002-3223-0754]{R.~Mukherjee}\affiliation{Department of Physics and Astronomy, Barnard College, Columbia University, NY 10027, USA}
\author[0000-0002-6121-3443]{W.~Ning}\affiliation{Department of Physics and Astronomy, University of California, Los Angeles, CA 90095, USA}
\author[0000-0002-9296-2981]{S.~O'Brien}\affiliation{Physics Department, McGill University, Montreal, QC H3A 2T8, Canada and Arthur B. McDonald Canadian Astroparticle Physics Research Institute, 64 Bader Lane, Queen's University, Kingston, ON Canada, K7L 3N6}
\author[0000-0002-4837-5253]{R.~A.~Ong}\affiliation{Department of Physics and Astronomy, University of California, Los Angeles, CA 90095, USA}
\author[0000-0001-7861-1707]{M.~Pohl}\affiliation{Institute of Physics and Astronomy, University of Potsdam, 14476 Potsdam-Golm, Germany and DESY, Platanenallee 6, 15738 Zeuthen, Germany}
\author[0000-0002-0529-1973]{E.~Pueschel}\affiliation{Fakult\"at f\"ur Physik \& Astronomie, Ruhr-Universit\"at Bochum, D-44780 Bochum, Germany}
\author[0000-0002-4855-2694]{J.~Quinn}\affiliation{School of Physics, University College Dublin, Belfield, Dublin 4, Ireland}
\author{P.~L.~Rabinowitz}\affiliation{Department of Physics, Washington University, St. Louis, MO 63130, USA}
\author[0000-0002-5351-3323]{K.~Ragan}\affiliation{Physics Department, McGill University, Montreal, QC H3A 2T8, Canada}
\author{P.~T.~Reynolds}\affiliation{Department of Physical Sciences, Munster Technological University, Bishopstown, Cork, T12 P928, Ireland}
\author[0000-0002-7523-7366]{D.~Ribeiro}\affiliation{School of Physics and Astronomy, University of Minnesota, Minneapolis, MN 55455, USA}
\author{E.~Roache}\affiliation{Center for Astrophysics $|$ Harvard \& Smithsonian, Cambridge, MA 02138, USA}
\author[0000-0003-1387-8915]{I.~Sadeh}\affiliation{DESY, Platanenallee 6, 15738 Zeuthen, Germany}
\author[0000-0002-3171-5039]{L.~Saha}\affiliation{Center for Astrophysics $|$ Harvard \& Smithsonian, Cambridge, MA 02138, USA}
\author{M.~Santander}\affiliation{Department of Physics and Astronomy, University of Alabama, Tuscaloosa, AL 35487, USA}
\author{G.~H.~Sembroski}\affiliation{Department of Physics and Astronomy, Purdue University, West Lafayette, IN 47907, USA}
\author[0000-0002-9856-989X]{R.~Shang}\affiliation{Department of Physics and Astronomy, Barnard College, Columbia University, NY 10027, USA}
\author[0000-0003-3407-9936]{M.~Splettstoesser}\affiliation{Santa Cruz Institute for Particle Physics and Department of Physics, University of California, Santa Cruz, CA 95064, USA}
\author{A.~K.~Talluri}\affiliation{School of Physics and Astronomy, University of Minnesota, Minneapolis, MN 55455, USA}
\author{J.~V.~Tucci}\affiliation{Department of Physics, Indiana University Indianapolis, Indianapolis, Indiana 46202, USA}
\author{V.~V.~Vassiliev}\affiliation{Department of Physics and Astronomy, University of California, Los Angeles, CA 90095, USA}
\author[0000-0003-2740-9714]{D.~A.~Williams}\affiliation{Santa Cruz Institute for Particle Physics and Department of Physics, University of California, Santa Cruz, CA 95064, USA}
\author[0000-0002-2730-2733]{S.~L.~Wong}\affiliation{Physics Department, McGill University, Montreal, QC H3A 2T8, Canada}
\author[0009-0001-6471-1405]{J.~Woo}\affiliation{Columbia Astrophysics Laboratory, Columbia University, New York, NY 10027, USA}

\collaboration{100}{(VERITAS Collaboration)}
\correspondingauthor{Lab Saha}
\email{lab.saha@cfa.harvard.edu}
\correspondingauthor{Wystan Benbow}
\email{wbenbow@cfa.harvard.edu}
\correspondingauthor{Martin Pohl}
\email{martin.pohl@uni-potsdam.de}

\begin{abstract}
Assuming Galactic cosmic rays originate in supernovae and the winds of massive stars, 
starburst galaxies should produce very-high-energy (VHE; E$>$100 GeV) gamma-ray emission 
via the interaction of their copious quantities of cosmic rays with the large reservoirs
of dense gas within the galaxies. Such VHE emission was detected by VERITAS from the starburst galaxy M\,82 in 2008-09.  
An extensive, multi-year campaign followed these initial observations, yielding a total of 254 h of good quality VERITAS data on M\,82.  Leveraging modern
analysis techniques and the larger exposure, these VERITAS data show a more statistically significant VHE signal ($\sim$6.5 standard deviations ($\sigma$)).  The corresponding photon spectrum is well fit by a power law
($\Gamma = 2.3 \pm 0.3_{stat} \pm0.2_{sys}$) and the observed integral flux is F($>$450 GeV) =
$(3.2 \pm0.6_{stat} \pm 0.6_{sys}) \times 10^{-13}~\mathrm{cm^{-2}~s}^{-1}$, or $\sim$0.4\% 
of the Crab Nebula flux above the same energy threshold. 
The improved VERITAS measurements, when combined with various multi-wavelength data, enable 
modeling of the underlying emission and transport processes. 
A purely leptonic scenario is found to be a poor representation of the gamma-ray spectral energy distribution (SED). 
A lepto-hadronic scenario with cosmic rays following a power-law spectrum in momentum  (index $s\simeq 2.25$), and with significant bremsstrahlung below $1$~GeV, provides a good match to the observed SED. The synchrotron emission from the secondary electrons indicates that 
efficient non-radiative losses of cosmic-ray electrons may be related to advective escape from the starburst core.
\end{abstract}



\section{Introduction}\label{sec:intro}
It is believed that Galactic cosmic rays (protons and nuclei) are dominantly accelerated by the winds and supernovae of massive stars \citep[][and references therein]{alma9924473798902466,Volk1996SSRv75279V,2009Natur.460..701B}. Accordingly, galaxies displaying an exceptionally high rate of star formation should harbor a correspondingly high cosmic-ray density. These cosmic rays  produce gamma-ray emission via their 
interaction with interstellar gas and radiation. 

M\,82 is a bright starburst galaxy located in the direction of the Ursa Major constellation at a distance of approximately 12 million light-years from Earth. It is an excellent laboratory for understanding the physics of star formation \citep{Sakai1999ApJ...526..599S}. An active starburst region in its center with a diameter of approximately 1000 light-years is home to hundreds of massive stars with a total stellar mass of about $10^4$ to $10^6 ~ \rm M_\sun$, where M$_\sun$ is the mass of the Sun \citep{Volk1996SSRv...75..279V,Melo2005ApJ...619..270M}. The galaxy has a star formation rate which is $\sim$10 times higher than galaxies such as the Milky Way, and its supernova rate is $\sim$0.1 to $\sim$0.3 per year. Most of these supernovae explode near the starburst core where the gas number density is around $200\ \mathrm{cm}^{-3}$ \citep{Kronberg1985ApJ...291..693K,Fenech2008MNRAS.391.1384F,Rieke1980ApJ...238...24R,2001A&A...365..571W}. The most recent Type Ia supernovae SN 2014j within M\,82 was detected on January 21, 2014 (MJD 56678) by the UCL Observatory \citep{Fossey2014CBET.3792....1F,Goobar2014ApJ...784L..12G}.

Starburst galaxies may be calorimeters for cosmic rays, at least for the electron component \citep{1989A&A...218...67V}, meaning that the particles lose their energy within the system and do not efficiently escape. For example, \citet{2013ApJ...768...53Y} calculate steady-state cosmic-ray spectra using a one-zone model assuming a single-value proxy for the varying environmental conditions in the multi-phase medium within M\,82 and find consistency with the calorimeter limit. The VERITAS data on M\,82 can further
test this hypothesis.

Given the high density of cosmic rays and ambient matter, M\,82 was long 
viewed as a promising target for detecting high-energy gamma rays. However, it was not detected above 100 MeV by the EGRET experiment \citep{1999ApJ...516..744B}, nor by any early ground-based gamma-ray instrument \citep{2005PhDT.........8N,gotting2006nachweis}. In 2009, the VERITAS collaboration reported the first detection of gamma rays ($>$ 700 GeV) from M\,82 \citep{2009Natur.462..770V}. Gamma rays were contemporaneously detected at MeV-GeV energies with 
the \textit{Fermi}-LAT telescope \citep{Abdo-2009}. These detections, 
together with the detection of gamma rays from the starburst galaxy NGC 253 by H.E.S.S. \citep{Acero_2009Sci...326.1080A} and \textit{Fermi}-LAT \citep{Abdo-2009}, and the \textit{Fermi}-LAT detection of star-forming region 30 Doradus in the Large Magellanic Cloud \citep{Abdo_2010} give strong support to the idea that supernovae and 
massive-star winds are dominant accelerators of cosmic rays up to energies of 
$\sim$$10^{15}$ eV. Hence, starburst galaxies are considered as a prominent, albeit rare, class of 
GeV-TeV gamma-ray emitter.

The initial detection of M\,82 is one of the weakest reported by VERITAS; the data yielded only $\sim$4.8$\sigma$ post-trials in 137 h, and the corresponding photon spectral index was not well constrained ($\Gamma = 2.5 \pm 0.6_{stat} \pm0.2_{sys}$). 
The dedicated long-term observations described here 
improve the photon spectrum (both in statistics and energy range) 
and correspondingly reduce the statistical uncertainty on the spectral index.

\section{VERITAS Observations and data reduction}\label{sec:observation}
VHE gamma-ray observations of M\,82 are performed using the VERITAS array of four 12-m diameter Imaging Atmospheric Cherenkov Telescopes (IACTs). VERITAS is located at the Fred Lawrence Whipple Observatory in Arizona, USA (31$^\circ 40'30''$ N, 110$^\circ57'07''$ W, 1268 m above the sea level). Each telescope is equipped with a camera of 499 photomultiplier tubes providing a field of view of about $3.5^\circ$. The typical energy threshold of the stereoscopic system is $\sim$100 GeV, and it provides an energy resolution of $\sim$15\% and angular resolution of $\sim$0.1$^{\circ}$ per event. VERITAS is able to detect a weak source ($\sim$0.6\% of the Crab Nebula flux) at 5$\sigma$ significance in 50 hours of observations at small ($<30^\circ$) zenith angles \citep{2015ICRC...34..771P,Adams2022A&A...658A..83A}.
   
\subsection{Data Analysis}
 
M\,82 was observed for a total of about 335 hours between 2007 and 2022 
at zenith angles between 37$^\circ$ and 50$^\circ$. The observations were performed in wobble-mode \citep{Fomin_1994} with the source
offset by $0.5^\circ$, alternating in each of the four cardinal directions,
to enable simultaneous background estimation.
After quality cuts, which account for hardware problems and 
poor atmospheric conditions, 254 hours of high-quality, 
four-telescope dark-time are selected for further analysis.
The analysis of these VERITAS data is performed using an up-to-date version of the standard VERITAS Analysis and Reconstruction Software  \citep[VEGAS 2.5.9;][]{2008ICRC....3.1385C}, following the standard analysis procedure. 
The data are reduced using the 
Image Template Method \citep[ITM;][]{2017ICRC...35..789C}, which provides
an improved angular resolution compared to early VERITAS publications.
The event-selection criteria for identifying
gamma-ray images and removing background cosmic-ray images
is optimized for hard-spectrum sources using Crab
Nebula data scaled to 1\% of its nominal strength.
The average energy threshold of this analysis is $\sim$450 GeV, and is
lower than the prior VERITAS publication on M\,82 due to
an upgrade of the VERITAS cameras completed in 2012 \citep{Kieda:2011iz,Otte:2011sm}.

The VERITAS signal is extracted from a circular region of $\sim$ 0.07$^\circ$ radius
centered on the position of M\,82. The significance of any excess is calculated following
Equation 17 of \citet{Li_Ma_1983}, where the background is estimated via a ring (inner radius $0.6^\circ$ and outer radius $0.8^\circ $) centered on the source position \citep[Ring Background Model;][]{Berge_et_all2007}.
As the ring model has an energy-dependent radial acceptance correction, 
the background is estimated differently for the spectral and flux measurements 
\citep[Reflected Region Method;][]{Berge_et_all2007}.  There are 
typically 15-18 non-overlapping regions, with the same offset ($0.5^\circ$) 
from the center of the VERITAS camera as the source which are used for the spectral analysis; 
some potential off-source regions are excluded due to bright stars. 
The VERITAS results were also independently verified using a second VERITAS analysis software package 
described in \citet{2017ICRC...35..747M}.

\begin{figure}[ht]
\centering
\includegraphics[width=\columnwidth]{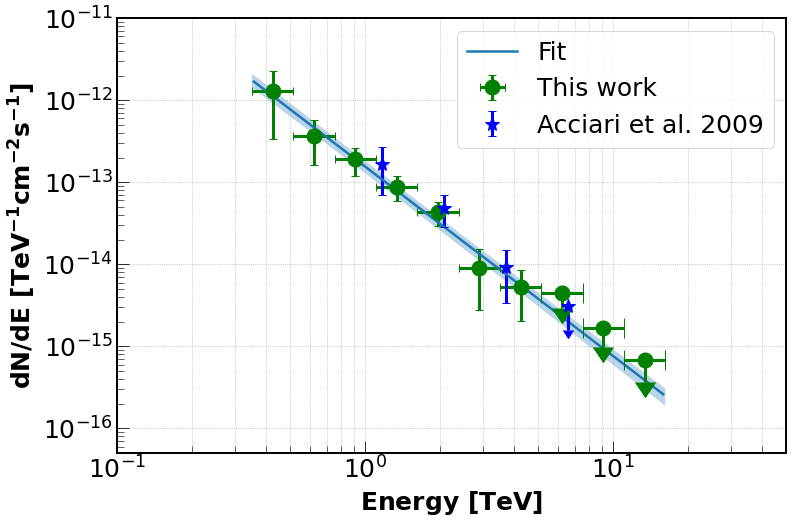}
\caption{VHE photon spectrum of M\,82 from the 2008-22 VERITAS observations. The 2008-09 result from the previous VERITAS publication  \citep{2009Natur.462..770V} is also shown. The 2008-22 data points are best fit by 
a power-law function ($\Gamma =  2.3$). The statistical uncertainty for the fit is also shown. \label{fig:gammaspec}}
\end{figure}

\subsection{Results}
A point-like excess of 135 gamma-ray-like events 
above the estimated background of 372 events
is observed from the direction of M\,82 in $254\,$h of lifetime after cuts. This excess corresponds to a statistical significance of 6.5 standard 
deviations ($\sigma$), or a chance probability of  $4.0 \times 10^{-11}$.
Table~\ref{ObservationSummary} shows the results from the analysis of the 
VERITAS data for a variety of epochs.  While the overall
significance is higher, the value observed for the original epoch (4.1$\sigma$) is lower than the previous
pre-trials value (5.0$\sigma$).  The clear detection from the latter, independent data set clearly 
confirms the previously reported VHE emission from M\,82. The time-averaged VHE photon spectrum measured
for the entire data set is shown in Fig. \ref{fig:gammaspec}.  These data are best fit by a 
power-law function with a photon index $ \Gamma = 2.3 \pm 0.3_{stat} \pm0.2_{sys}$, with flux normalization     
(7.2 $\pm 1.2_{stat}  \pm 1.4_{sys}) \times 10^{-14}$ cm$^{-2}$ s$^{-1}$ at 1.4 TeV,
in the energy range from 200 GeV to 5 TeV. The $\chi^2$ of the power-law fit is 1.4 for 5 degrees of freedom, 
corresponding to a probability of 0.92. The observed photon index is in agreement with the early 
VERITAS result (cf. $\Gamma = 2.5 \pm0.6_{stat} \pm 0.2_{sys}$),. There is no evidence for a high-energy cut-off in the spectrum,  and Sec. ~\ref{sec:model} includes brief discussion on what this implies for the spectrum of the radiating particles. 

\begin{table*}[ht]
\centering
\begin{tabular}{c | c c c c c c | c c | c }
\hline\hline
Epoch & T & On & Off & $\alpha$ & Excess & $\sigma$ & F$(> 450$ GeV) & Crab & F$(> 700$ GeV) \\
& [h] & & & & & & [ $10^{-13}~\mathrm{cm^{-2}~s}^{-1}$ ] & [ \% ] & [ $10^{-13}~\mathrm{cm^{-2}~s}^{-1}$ ]\\
\hline
2008-09 &137 & 220 & 7106 & 0.02316 & 55.4 & 4.1 &$3.4 \pm 1.0_{stat} \pm 0.8_{sys}$ & 0.4 & $ 1.9\pm 0.6_{stat} \pm 0.4_{sys}$\\
\hline
2011-22 &117 & 286 & 8849 & 0.02346 & 79.6 &  5.2 & $3.2 \pm 0.8_{stat} \pm 0.6_{sys}$ & 0.4 & $ 1.8\pm 0.5_{stat} \pm 0.4_{sys}$  \\
\hline
Total (2008-22) & 254 & 506 & 15955 & 0.02330 & 135.0 & 6.5 &$3.2 \pm 0.6_{stat} \pm 0.6_{sys}$ & 0.4 &$1.8 \pm 0.3_{stat} \pm 0.4_{sys}$ \\
\hline\hline
\end{tabular}
\caption{Results from the analysis of VERITAS data on M\,82. The top row (2008-09) is the updated analysis from the epoch originally published in  \citep{2009Natur.462..770V}. The second row is for all data since the original publication.
The bottom row is for the entire data set.  The quality-selected live time, number of $\gamma$-ray-like events in the on- and off-source regions, the normalization for the larger off-source region, the observed excess of $\gamma$-rays and the corresponding statistical significance are shown. The flux is reported above the observation threshold of 450 GeV, and is also given in percentage of Crab Nebula flux above the same threshold. For comparison to the flux from the original VERITAS publication,
the corresponding flux extrapolated above the higher original threshold of 700 GeV is also given.}
\label{ObservationSummary}
\end{table*}

The observed gamma-ray flux from M\,82 is F($>$450 GeV) = $(3.2 \pm 0.6_{stat} \pm 0.6_{sys}) \times 10^{-13}~\mathrm{cm^{-2}~s}^{-1}$. This corresponds to $\sim$0.4\% of the Crab Nebula flux above the same threshold.   Extrapolating this flux to
the threshold of the original VERITAS publication \citep{2009Natur.462..770V} yields F($>$ 700 GeV) = $(1.8 \pm 0.3_{stat} \pm 0.4_{sys}) \times 10^{-13}~\mathrm{cm^{-2}~s}^{-1}$, which is approximately half of the original 
value (F($>$ 700 GeV) = $(3.7 \pm0.8_{stat} \pm 0.7_{sys}) \times 10^{-13}~\mathrm{cm^{-2}~s}^{-1}$).
A re-analysis of the original sample (2008-09 data) with the improved calibration, understanding of the detector and simulations, as well as with the more sensitive ITM analysis technique \citep{Jodi_Christiansen_2018} yields a reduced significance, and also a decreased flux
(F($>$ 700 GeV) = $(1.9 \pm 0.6_{stat} \pm 0.4_{sys}) \times 10^{-13}~\mathrm{cm^{-2}~s}^{-1}$).
While the results remain in reasonable agreement given the large errors on the early measurement, 
this suggests that the flux difference primarily comes from an improved understanding of the instrument and 
somewhat from statistical fluctuations in the background, rather than a change in the source over time.

\begin{figure}[ht]
\centering
\includegraphics[width=\columnwidth]{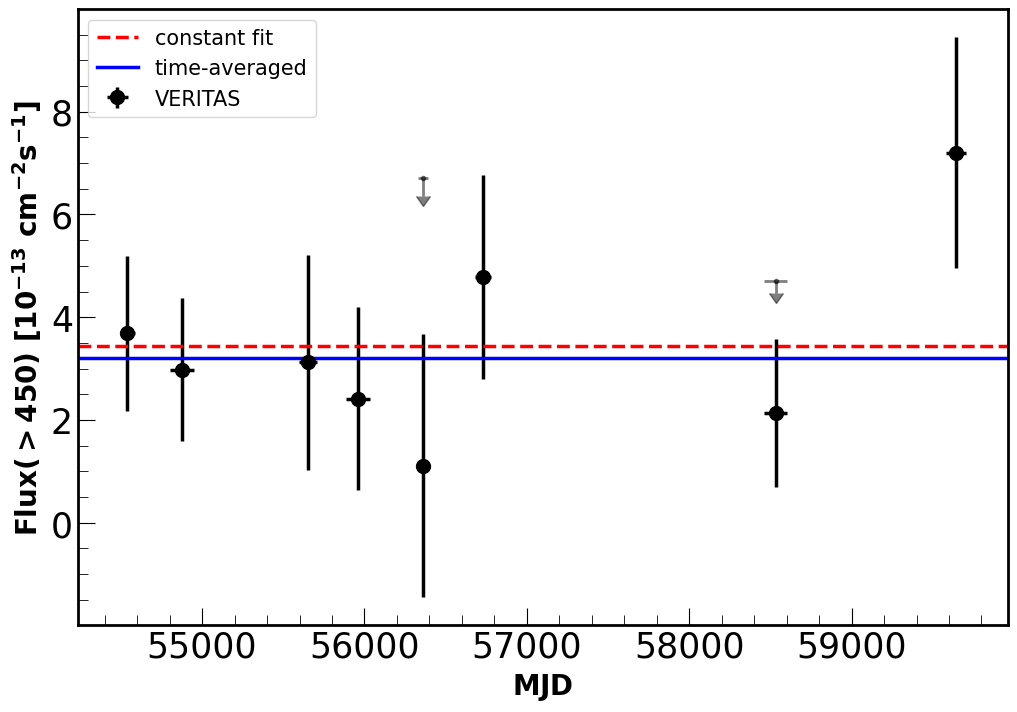}
\caption{The integral flux above 450 GeV observed by VERITAS during each season. Upper limits (99\% confidence level) are also shown for the 2012-13 and 2018-19 seasons due to the low significance observed. 
The time-averaged flux measured during the entire observation period (2007 to 2022) is indicated by a solid (blue) line. 
A fit of a constant to the seasonal data is shown by the dashed (red) line.\label{fig:lc}}
\end{figure}

To better illustrate the steady flux observed from M\,82, the VHE light curve is shown in Figure \ref{fig:lc}.
Here the integral flux above 450 GeV measured for each season (September to July) is plotted.
A fit of a constant function to these data ($\chi^2$=5.4 for 7 degrees of freedom, P($\chi^2$)= 0.61), as well as the time-average 2008 - 2022 flux,  are also shown.  There is no evidence of variability in the seasonal flux measurements.

\begin{figure*}[ht]
\centering
\includegraphics[width=0.6\textwidth]{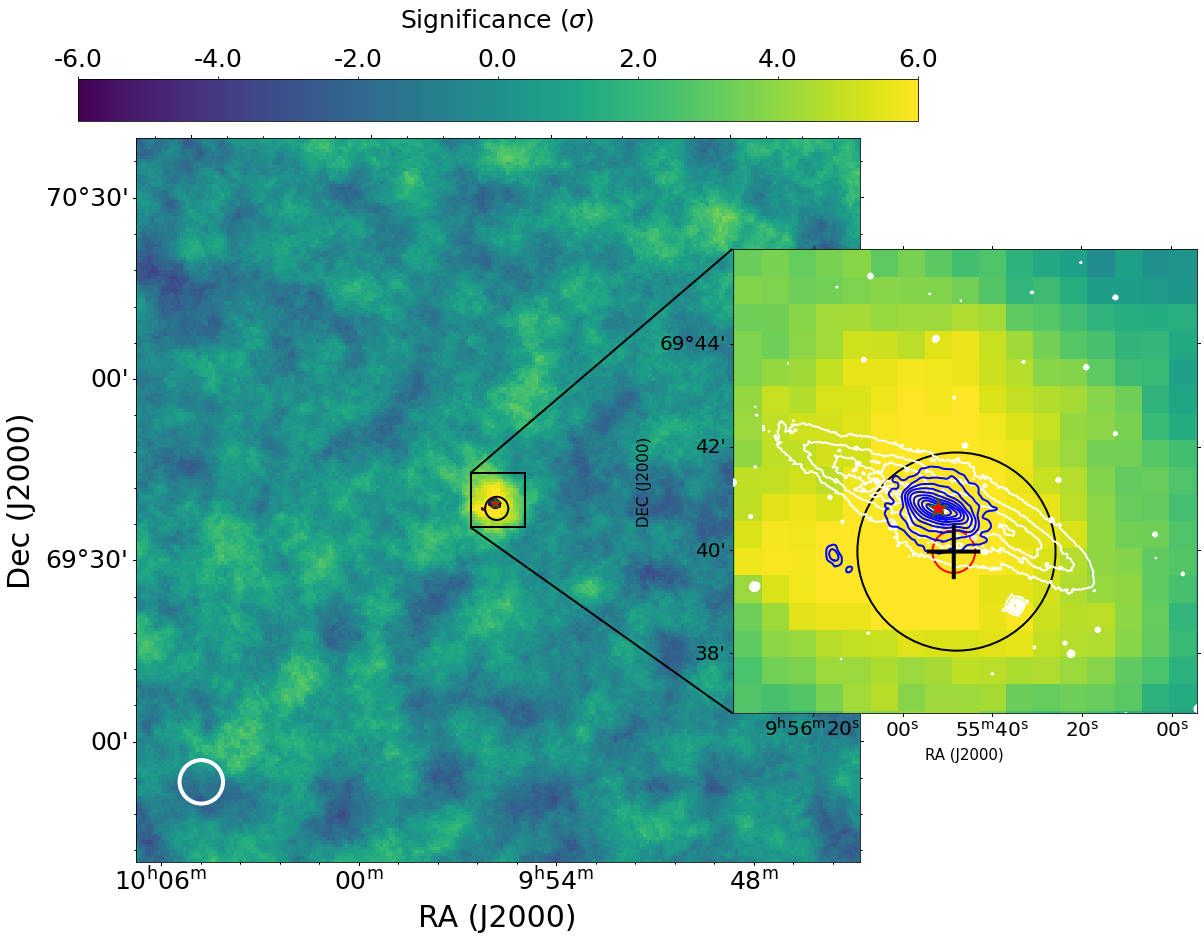}
\caption{Sky map of significance observed by VERITAS from the region near M\,82. The position of
M\,82 is indicated with a red star \citep{2007MNRAS.376..371J}. The total power radio continuum contours (blue) at about 10 GHz from the WSRT observations are overlaid on the skymap \citep[see Fig. 3 of ][]{Adebahr_2013A&A...555A..23A}. These radio contours start at the 3$\sigma$ level of the 50 $\mu$Jy/beam, with a beam size of $7''.8$ $\times$ $7''.3$, and increases in powers of 2. The data for optical contours (white) at g-band are taken from the survey data of stellar structure in galaxies by \textit{Spitzer} \citep{Knapen2014A&A...569A..91K}. The optical contour starts at 16 magnitude and goes to 1 in step 2.143. The black circle indicates the upper limit of the extension (1.9$'$) of M\,82 as discussed in section \ref{sec:observation}. The best-fit location with 1$\sigma$ statistical errors is shown with a black cross. The 1$\sigma$ systematic uncertainties are shown around the best-fit centroid position with a red circle. The white circle indicates the VERITAS point spread function 
(68\% containment radius) at the average energy threshold of $\sim$450 GeV. \label{fig:skymap}}
\end{figure*}

The best-fit location of the source and its extension 
are estimated using the methodology discussed in \citet[][and references therein]{2020ApJ...894...51A}. 
To estimate the Point Spread Function (PSF) of VERITAS, Crab Nebula data are analyzed considering VERITAS data from
similar epochs and zenith angle coverage to M\,82. Although a slight extension ($\sim$1 arcmin) of the Crab Nebula was discovered at GeV--TeV energies \citep{HESS_2020NatAs...4..167H,HESS_2024A&A...686A.308A}, this small effect is not accounted for in this analysis. The PSF is modeled using a two-dimensional 
Moffat function, as described by the following equation:
\begin{equation}
K(x,y) = A \Big[ 1 + \frac{(x-x_{0})^{2} + (y-y_{0})^{2}}{\gamma ^{2}}\Big]^{-\alpha}
\label{2dking}
\end{equation}
Here, $x_{0}$ and $y_{0}$ denote the positions of the maximum of the Moffat model in the x and y directions, respectively. 
$\gamma$ and $\alpha$ represent the core width and power index of the Moffat model. The values of $\alpha$, $\gamma$ 
are estimated to be 2.41 and 0.069 respectively. The best-fit for the location (J2000) of the peak of the VERITAS gamma-ray emission is given 
by RA=$9^h55^m48.6^s  \pm 5.5^s_{\mathrm stat} \pm 4.8^s_{\mathrm sys}$  and DEC =$+69^{\circ} 39' 58.5'' \pm 28.8''_{\mathrm stat} \pm 25''_{\mathrm sys}$. 
It is consistent with the position of M\,82 as given in \citet{2007MNRAS.376..371J} (i.e. within an angular separation of $0.014^\circ$), 
and is named VER\,J0955+696.  It is also evident from the sky map of the region surrounding M\,82 (see Fig. \ref{fig:skymap}) with the PSF at the bottom left corner, that the VERITAS excess is point-like at GeV-TeV energies. The upper limit (UL) on the extension of the VHE excess from M\,82 is determined 
to be 1.9 arcminutes at 95\% confidence level. 

The best-fit location of the emission observed by VERITAS from M\,82 and the core position of the Galaxy are shown in Fig. \ref{fig:skymap}. 
This best-fit location is $\simeq0.9\sigma$ away from the core position of M\,82 when both the systematic and statistical errors are considered,
suggesting that these two locations are consistent with each other.  Significantly more data would be needed to try to distinguish whether the
origin of the gamma-ray emission differs from the core of the galaxy based on positional information alone.

\section{Multi-wavelength data} \label{sec:mw-data}

In addition to VERITAS data, archival fluxes observed in different wavebands, 
from radio through high-energy gamma rays, are considered for 
the modelling of M\,82. Only spectral data points resulting from the non-thermal emission processes are considered.
Brief details of these observations are discussed below. 

\subsection{Radio observations}\label{sub:radio}
\begin{figure*}[ht!]
\centering
\includegraphics[width=0.8\textwidth]{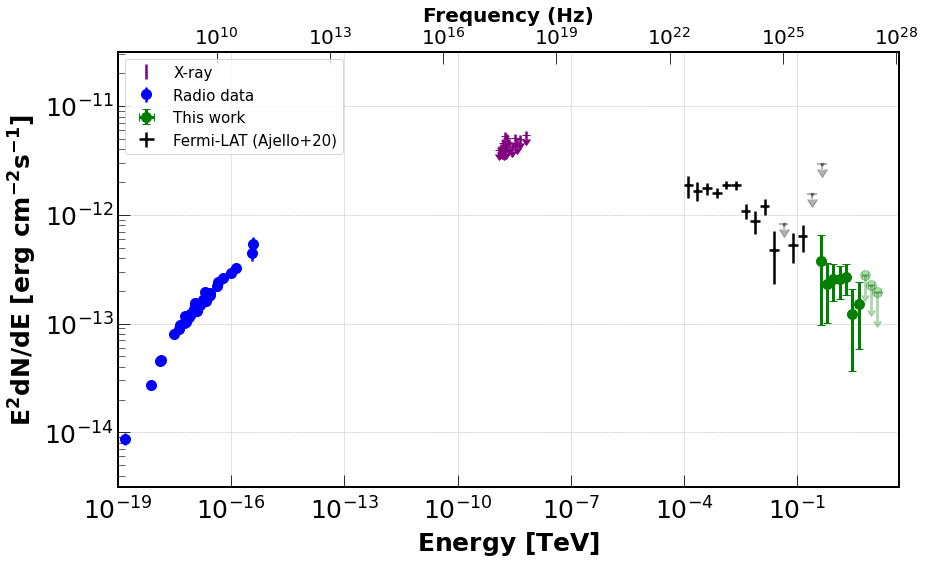}
\caption{The spectral energy distribution (SED) of M\,82 from radio to TeV energies. The radio data are compiled from 
observations discussed in section \ref{sub:radio} and are shown with a single marker (circle) and color (blue). The X-ray data 
in the energy range of 1.25 -- 6.5 keV are shown in line (purple) and taken from \citet{Brightman_2020ApJ...889...71B}. 
The \textit{Fermi}-LAT data are taken from \cite{Ajello_2020ApJ...894...88A}. The VERITAS data are the results 
presented in this paper.\label{fig:sed}}
\end{figure*}

The first radio observation of M\,82 was conducted more than 30 years ago by \citet{Condon_1992ARA&A..30..575C}. 
Later, M\,82 was imaged at 327 MHz with 40$''$ resolution using the Westerbork synthesis radio telescope (WSRT) by
\citet{Adebahr_2013A&A...555A..23A}. These observations established that M\,82 has a radio-bright extended halo structure. 
Figure~\ref{fig:sed} displays the radio spectrum above $0.8$~GHz 
\citep{1988A&A...190...41K,1991ApJ...366..422C,2010ApJ...710.1462W,2012MNRAS.419.1136B,Adebahr_2013A&A...555A..23A}. 
At frequencies below $\sim$1 GHz, free-free absorption is expected to modify the spectrum \citep{1994A&A...287..453P}, and data in this frequency band are not considered.

\subsection{X-ray observations}
M\,82 is extensively studied in the X-ray band. Using data from {\it Chandra} and {\it NuSTAR} shows 
that M\,82 hosts two ultraluminous X-ray (ULXs) sources: X-1, an intermediate-mass black hole candidate, 
and X-2, an ultraluminous X-ray pulsar \citep{Matsumoto2001ApJ...547L..25M,Bachetti2014Natur.514..202B}. While {\it Chandra} observations provide high spatial
resolution in the 0.5--8 keV band, {\it NuSTAR} provides sensitive spectral coverage above 10 keV. 
\citet{Brightman_2020ApJ...889...71B} studied the spectral evolution of X-1 and X-2 finding that the total X-ray emission from M\,82 is highly variable, most likely due to X-1. For the multi-waveband 
spectral modelling, only the diffuse emission components in the energy range above $\sim$1.25 keV to 8 keV 
are considered, and the flux attributable to the compact sources is subtracted. 
Since the diffuse emission could still be a mix of other compact sources, thermal emission, and a
nonthermal component, the diffuse flux component is treated as an upper limit for constraining the emission models.  

\subsection{MeV-GeV gamma-ray observations}
Using roughly 10 years of gamma-ray data taken by the Fermi Large Area Telescope \citep{Ajello_2020ApJ...894...88A},
M\,82 is studied in the energy range E=[0.1, 800] GeV. 
The starburst galaxy is detected with a significance of 34.4$\sigma$, corresponding to an integral flux 
above 1 GeV of $(1.13 \pm 0.05)~\times 10^{-11}~\mathrm{cm^{-2} s}^{-1}$. 
The photon spectrum is best fit by a power law with index $\Gamma = 2.14 \pm 0.06$, 
which is consistent with the spectral shape in the VHE band. No variability is observed 
in the 10 years of \textit{Fermi}-LAT observations from M\,82. 

\section{Modelling the Spectral Energy Distribution}\label{sec:model}

The observed emission can be interpreted with leptonic and hadronic emission models. Early assessments, 
prior to the gamma-ray detection of M\,82, predicted a GeV-scale flux a factor of a few below the 
sensitivity limit of EGRET \citep{1994A&A...287..453P,2008A&A...486..143P}.
An important finding was that the radio-synchrotron emission from secondary electrons produced in the same interactions that 
create hadronic gamma rays should be particularly bright, on account of the rapid energy losses of electrons
and the typical duration of starburst events. Therefore the hadronic gamma-ray output could 
not be arbitrarily bright, and it would always be accompanied by leptonic radiation in the radio and in X-ray
bands.

\subsection{Leptonic scenario}\label{lep-sce}
Within a purely leptonic scenario the emission from electrons has to account for the entire
non-thermal SED, in particular the bright GeV-scale emission with flux 
$\nu F_\nu \approx 10^{-12}\ \mathrm{erg\,cm}^{-2}\,\mathrm{s}^{-1}$. 
There are two relevant radiation processes in this energy band, non-thermal bremsstrahlung and 
inverse-Compton scattering \citep{1970RvMP...42..237B}. 

The interstellar medium (ISM) in M\,82 is composed of very dense molecular clouds, dense warm, ionized gas, and a hot, low-density medium in approximate pressure balance \citep{Westmoquette_2009}. One can simplify the complex structure as a uniform zone with ``mean" values for the environmental parameters, as was done in other studies \citep[e.g.][]{1994A&A...287..453P,2008A&A...486..143P,deCeadelPozo_2009,Paglione_2012,2013ApJ...768...53Y,Lacki_2013}. Among these studies, there is a range of ``mean" parameter values that were adopted. The fiducial values used here are in the middle of this range and are listed in Table~\ref{tab2}. To reflect the impact of the uncertainty in the parameters, one can add scaling factors to the equations where applicable. In any case, the ratios of the parameter values display considerably less variation than do the values themselves, which renders estimates of the energy partition between the various radiation and energy-loss channels reasonably well defined. 

Relativistic electrons are known to quickly lose energy through various channels, on a time scale that is shorter than the duration of the starburst in M82 \citep{2003ApJ...599..193F}. The total energy loss rate, $b(E)$, can be written as \citep{1993A&A...270...91P}
\begin{equation}
-b(E) =C_1+C_2\,E + C_3\,E^2 
\label{eq7}
\end{equation}
where
\begin{align}
C_1=& \left(3.7\cdot 10^{-16}\ \mathrm{GeV\,s}^{-1}\right)\,\left(n_H+1.54\,n_e\right) \nonumber \\
C_2=& \left(10^{-15}\ \mathrm{s}^{-1}\right)\,\left(n_H+0.95\,n_e\right)  \nonumber \\
C_3=& \left(10^{-16}\ \mathrm{GeV}^{-1}\,\mathrm{s}^{-1}\right)\,\left(U_\mathrm{mag}+U_\mathrm{rad}\right) \ .\nonumber 
\label{eq7g}
\end{align}
This includes the loss rates for ionization and Coulomb scattering ($C_1$), nonthermal bremsstrahlung ($C_2$), and synchrotron emission and inverse-Compton radiation ($C_3$).
The energy densities of the magnetic field and the soft radiation, $U_\mathrm{mag}$ and $U_\mathrm{rad}$, are in units of $\mathrm{eV\,cm}^{-3}$, and the number densities of atomic ($n_H$) and ionized ($n_e$) gas are in cgs units. For simplicity, inverse-Compton scattering in the Thomson limit is assumed; this should be well justified for the production of GeV-scale gamma rays. 

If the rate of cosmic-ray production during the starburst phase in M\,82 was high for more than a few hundred thousand years, then the system would be calorimetric for electrons at energies around a GeV and higher \citep{1989A&A...218...67V}. In this calorimetric limit, a spectral break in the energy-loss rate imposes a corresponding feature in the electron spectrum,
\begin{equation}
N(E)=\frac{1}{\vert b(E)\vert}\int_E dE'\ Q(E')\ ,
\label{eq-ss}
\end{equation}
where $Q(E)$ is the source spectrum of electrons. A first change in the energy dependence of the loss rate, $E/\vert b(E)\vert$, is observed at about $E_{c1} =C_1/C_2\simeq 0.4$~GeV, which for $U_\mathrm{mag}\approx 500$ would appear at the same photon energy ($1.5\cdot 10^{-18}$~TeV or $400~$MHz) at which free-free absorption may become significant \citep{1994A&A...287..453P}. One can ignore synchrotron photon energies below $1~$GHz and therefore neglect the term $C_1$ in the loss rate. 

A second spectral feature should appear at
\begin{equation}
E_{c2}=\frac{C_2}{C_3}\simeq (2.5\ \mathrm{GeV}) \frac{1000 (n_H+0.95\,n_e)}{245(U_\mathrm{mag}+U_\mathrm{rad})} ,
\label{eqec2}
\end{equation}
where the number and energy densities are in cm$^{-3}$ and eV~cm$^{-3}$, respectively, and the numerical factors derive from the fiducial values of the parameters (cf. Table~\ref{tab2}). The spectral break should be seen in the radio spectrum at the frequency
\begin{equation}
\nu_{c2}\simeq (15\,\mathrm{GHz})
\sqrt{\frac{U_\mathrm{mag}}{500}}\left( \frac{1000 (n_H+0.95\,n_e)}{245(U_\mathrm{mag}+U_\mathrm{rad})}\right)^2 .
\label{eq7h}
\end{equation}
The frequency 
$15~$GHz corresponds to a photon energy of $6\cdot 10^{-17}~$TeV. No evidence of a high-frequency break in the radio synchrotron spectrum is seen, although it should be there, because the lifetime of the radiating electrons ($<$$10^5$~yrs) is very much shorter than the starburst duration and the evolutionary timescale of massive stars. 

The absence of a visible spectral break suggests the scaling factor might be larger than unity. However, the flux points above $20~$GHz ($10^{-16}$~TeV) could include dust emission or thermal bremsstrahlung, in which case the scaling factor for the energy-loss break could be close to unity. The spectral break near $E_\gamma\approx  10^{-18}$~TeV ($200~$MHz) in the radio spectrum may reflect free-free absorption. 
Indeed, homogeneously distributed ionized gas at $3000~$K temperature, that subtends a solid angle of $\Delta\Omega\approx 3\cdot 10^{-8}\,$sr and provides an opacity of unity at $400~$MHz, would provide a flux at $100~$GHz of \citep[sect. 9.4]{2004tra..book.....R}
\begin{equation}
E^2 \frac{dN}{dE}_\mathrm{free-free}(100\,\mathrm{GHz})\approx
3\cdot 10^{-13 }\ \mathrm{erg/cm^2/s} ,
\label{eqff}
\end{equation}
which is nearly the entire observed flux. It is therefore likely that in the band $10$~GHz to $100~$GHz the transition from synchrotron-dominated emission to thermal radiation is seen. 

\begin{figure}[ht!]
\centering
\includegraphics[width=\columnwidth]{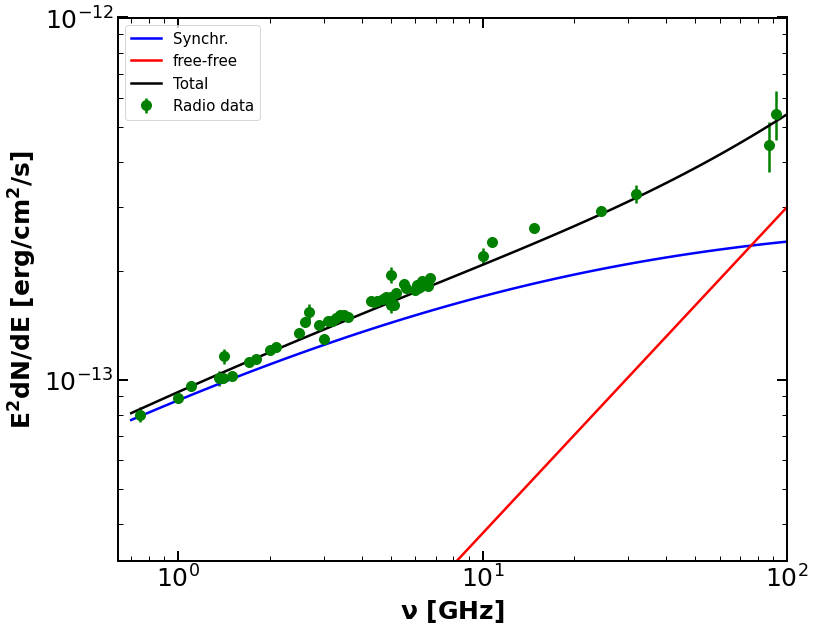}
\caption{Radio SED of M\,82 compared with a model spectrum that is composed of a synchrotron component following equation~\ref{eq-ss} with $Q\propto E^{-2.25}$ and free-free emission with a flux slightly below the estimate in equation~\ref{eqff}.
The radio data are taken from \citet{1988A&A...190...41K,1991ApJ...366..422C,2010ApJ...710.1462W,2012MNRAS.419.1136B,Adebahr_2013A&A...555A..23A}.
\label{fig:radio-model-brems}}
\end{figure}

Figure~\ref{fig:radio-model-brems} displays the radio spectrum in comparison with a free-free emission component and a synchrotron component for a scaling factor of unity (see equation \ref{eq7h}) and a spectrum according to equation~\ref{eq-ss} with $Q\propto E^{-2.25}$. The reasonable match between the model and data is notable, although the parameters have only been coarsely optimized.
 
There will be a bremsstrahlung contribution to the GeV-scale gamma-ray emission, and, using the differential cross section for bremsstrahlung given in \citet{1970RvMP...42..237B} and our fiducial parameters, the observed radio synchrotron spectrum in the frequency band $1$~GHz to $10$~GHz can be turned into an estimate for the bremsstrahlung spectrum that is displayed in Figure~\ref{fig:model}. More specifically, the fraction of the observed flux at $E=0.15$~GeV provided by bremsstrahlung is
\begin{equation}
\frac{F_\mathrm{bre}(0.15\,\mathrm{GeV})}{F_\mathrm{obs}(0.15\,\mathrm{GeV})}\simeq 0.8\ 
\frac{n_H+0.95 n_e}{245} \left(\frac{500}{U_\mathrm{mag}}\right)^{5/6} , 
\label{eq_scalebr}
\end{equation}
where as in Eq.~\ref{eqec2} the number and energy densities are in cm$^{-3}$ and eV~cm$^{-3}$, respectively, and the numerical factors derive from the fiducial values of the parameters. 
This fraction of flux rapidly falls off with energy, and for the fiducial parameter values corresponds to only 30\% at $1$~GeV. To be noted from Eq.~\ref{eq_scalebr} is that the parameter scaling factor for the bremsstrahlung flux cannot be much larger than unity, otherwise the bremsstrahlung flux would overshoot the observed gamma-ray flux near $0.1$~GeV.

\begin{table}
\begin{center}
\begin{tabular}{ l| r l }
\hline
 \hline
 Parameter& Value & Unit\\
 \hline
 $n_H$ & 200 & cm$^{-3}$ \\
 $n_e$ & 50 & cm$^{-3}$ \\
 $U_\mathrm{mag}$ & 500 & eV\,cm$^{-3}$ \\
 $U_\mathrm{rad}$ & 500 & eV\,cm$^{-3}$ \\
   \hline
\end{tabular}
\caption{Fiducial parameter values for our radiation modeling. From top to bottom, we list the density of neutral \citep[cf.][]{1991MNRAS.252P...6S,2001A&A...365..571W,2010ApJ...722..668N} and ionized gas, and also the energy density of the magnetic field, here estimated above cosmic-ray equipartition on account of turbulent amplification \citep{2007mhet.book...85S,2009ApJ...693.1449C}, and the ambient infrared radiation \citep{1994MNRAS.270..641H,2003ApJ...599..193F}.}\label{tab2}
\end{center}
\end{table}
If the observed gamma-ray emission near $10~$GeV were the result of inverse-Compton scattering of the FIR radiation field \citep{2003ApJ...599..193F}, then the radiating electrons would have an energy of about 
$500~$GeV. This is high enough for the electrons to be efficiently cooled, and the spectral index should be softer by $1/2$ than that of the radio-synchrotron emission, implying a spectral index $\alpha=-2.13$ in $dN/dE$. The observed gamma-ray spectrum has a similar shape up to about a TeV, where the maximum electron energy and the Klein-Nishina transition would introduce a decline of flux. 

The synchrotron radiation from these $500$-GeV electrons would be observed in the optical, at $1$~eV or $10^{-12}$~TeV, where starlight dominates. As Thomson scattering off the FIR photons applies and the energy-loss rates for synchrotron and inverse-Compton emission are similar, so are the $\nu F_\nu$ flux and the spectral index at corresponding energies, $10~$GeV for inverse-Compton emission and $10^{-12}$~TeV for the synchrotron radiation. The Klein-Nishina transition is expected to be slightly higher in gamma-ray energy, near $1$~TeV. Extrapolating the synchrotron spectrum to $ 10^{-12}$~TeV (cf. Fig.~\ref{fig:radio-model-brems}) and using the fiducial ratio $U_\mathrm{rad}/U_\mathrm{mag}=1$, we estimate that inverse-Compton scattering only provides about $10\%$ of the flux at $10~$GeV. 

A significant contribution to the GeV-to-TeV gamma-ray emission from M\,82 would require the ratio $U_\mathrm{rad}/U_\mathrm{mag}$ to be much larger than unity. The FIR intensity distribution is well measured, and hence $U_\mathrm{rad}$ is well determined, but there is uncertainty in the energy density of the magnetic field. If it were smaller than $500\,$eV~cm$^{-3}$ by a certain factor, then the gas density in the starburst core would need to be reduced in a similar way, otherwise the bremsstrahlung flux at a few hundred MeV would overshoot the observed flux (cf. Eq.~\ref{eq_scalebr}). Indeed, the bremsstrahlung flux needs to significantly undershoot the observed flux, because unlike for the hadronic-emission scenario the inverse-Compton component provides approximately the same fraction of the $100$-MeV flux as it does to the $10$-GeV flux.
The modifications of the gas density and the magnetic-field strength drive the break frequency in the radio spectrum (Eq.~\ref{eq7h}) well below $1$~GHz. This would need to be compensated with a harder production spectrum of electrons, otherwise the observed radio spectrum would be poorly reproduced.  Reducing the energy density of the magnetic field by a factor three, and likewise the gas density, already requires a source spectrum $Q\propto E^{-1.9}$, and the inverse-Compton spectrum would no longer match the observed gamma-ray spectrum.

We conclude that a purely leptonic scenario for the SED of M\,82 is very unlikely, although bremsstrahlung may provide a significant contribution to the observed gamma-ray flux below $1~$GeV. If that is the case, the electron-to-ion ratio of cosmic rays in M\,82 would be unusually large with $10\%$ or more at a few GeV.

\subsection{Hadronic scenario}\label{had-sce}
The winds of massive stars in the starburst region, 
as well as the resulting supernova explosions, 
will enrich the gas with metals that likely lead to a heavy composition of the cosmic-ray hadrons. 
By analyzing the gamma-ray SED, the spectrum of cosmic-ray nuclei and their composition can potentially be extracted, assuming they account for most of the measured GeV/TeV flux.

Here, $\eta_A$ is denoted as the fraction in number of a certain element with mass number $A$. It is simple to apply this factor to the target material, but it must also be accounted for in the cosmic-ray spectra. For calculation of the gamma-ray spectra, a model derived with the Monte-Carlo event generator DPMJET III is used; this model is described in \citet{2020APh...12302490B}. The emissivities are binned for cosmic-ray spectra written in total energy per nucleon, $E_A$. The actual particle acceleration scales with momentum per charge, known as the rigidity, $r=p/Z$. All cosmic rays are most likely injected at the same rigidity, because of a requirement 
that particles can cross the shock without significant deflection, and their spectrum terminates at the same maximum rigidity.

The energy per nucleon and the rigidity are related as
\begin{equation}
E_A=\sqrt{m^2c^4 +\frac{Z^2}{A^2} r^2c^2}
\label{eq1}
\end{equation}
where $m$ denotes the proton rest mass and $c$ is the speed of light.
It follows that the maximum particle energy scales as 
\begin{equation}
E_{A,\mathrm{max}}\propto \frac{Z}{A} .
\label{eq3}
\end{equation}
The injection rigidity of all particles being the same, the injection energy varies with mass and charge number, following Eq.~\ref{eq1}.
The modeling ignores that heavy elements may initially be locked up in dust particles \citep[e.g.][]{1997ApJ...487..182M,1997ApJ...487..197E}. For a power-law spectrum in rigidity with index $s$ and step functions for the smallest ($r_\mathrm{inj}$) and the largest rigidity ($r_\mathrm{max}$),
\begin{equation}
N(r)=N_0\ r^{-s}\,\Theta(r-r_\mathrm{inj})\,\Theta(r_\mathrm{max}-r) ,
\label{eq4}
\end{equation}
the corresponding spectrum in total energy per nucleon is
\begin{align}
N(E_A)=&\eta_A\,N_0\,E_A\,\left(\frac{A}{Z}\right)^{1-s}\,\left(E_A^2-m^2c^4\right)^{-(1+s)/2}\nonumber \\
&\times \,
\Theta\left(E_A-E_{A,\mathrm{inj}}\right)\,\Theta\left(E_{A,\mathrm{max}}-E_A\right) .
\label{eq5}
\end{align}
The overall normalization, $N_0$, should be the same for all elements, because the abundances are accounted for with the factor $\eta_A$. 

This model is considered with the elements hydrogen, helium, carbon, and oxygen matching the ISM composition and, as a test case,  with the number fractions matching 40\% ISM, 40\% Red-Supergiant winds, and 10\% each for WN and WC Wolf-Rayet stars. The composition fractions are applied to both the cosmic rays and the target gas. All cosmic rays are assumed to be fully ionized.

\begin{figure}[ht!]
\centering
\includegraphics[width=\columnwidth]{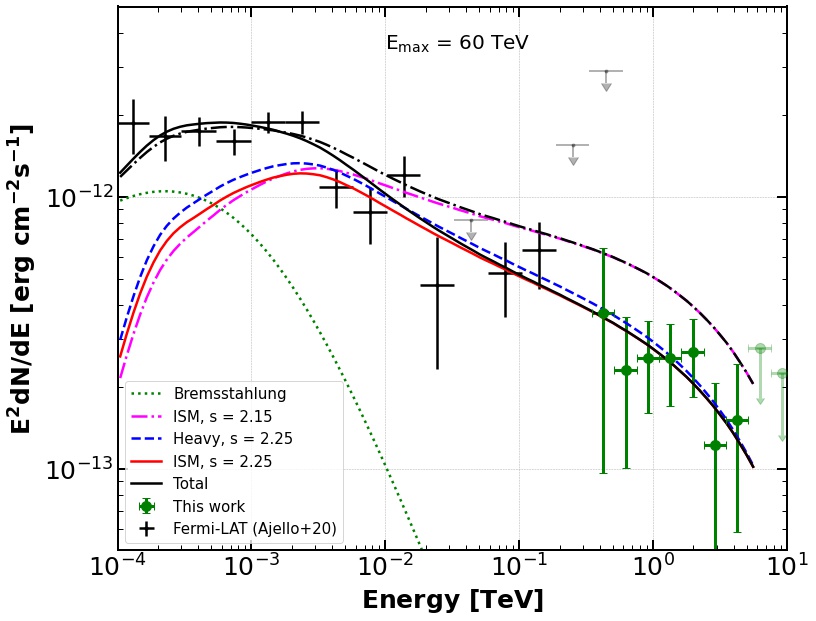}
\caption{Gamma-ray SED of M\,82 compared with hadronic models, varying the spectral index and the elemental composition. The bremsstrahlung contribution is given by the green dotted line. The maximum particle energy, $E_\mathrm{max}$, is set to $60$~TeV. The black solid line is the sum of bremsstrahlung and the hadronic component for ISM and $s=2.25$, and likewise the black dash-dotted line for $s=2.15$.}\label{fig:model}
\end{figure}

\begin{table}
\begin{center}
\begin{tabular}{ l| c c  }
\hline
 \hline
 Components& ISM&  Heavy\\
 \hline
 Hydrogen & 0.909 & 0.848 \\
 Helium & 0.090  & 0.146  \\
 Carbon & 2.1e-4 & 5.2e-3 \\
 Oxygen & 1.6e-4 & 7.e-4 \\
  \hline
\end{tabular}
\caption{Number fractions of the elements considered in the two models of composition, following \citet{2012A&A...540A.144S,2015A&A...579A..75T,2017A&A...605A..83D,2019A&A...621A..92S}}.\label{tab1}
\end{center}
\end{table}
Figure~\ref{fig:model} displays the gamma-ray SED in comparison with three model spectra and the bremsstrahlung component derived in section~\ref{lep-sce}. The normalization of the models has been determined by eye-adjustment only, on account of the systematic uncertainties in the bremsstrahlung contribution and emission from outside of the starburst core. To be noted from the figure are three points:
\begin{enumerate}
\item Whereas the spectrum above a few GeV appears to be well represented by a power law, the maximum proton energy, $E_\mathrm{max}$, is poorly constrained and may be below $60$~TeV. 
\item A heavy composition provides additional gamma-ray flux below a few GeV, but the uncertainties of the spectral data and the bremsstrahlung contribution are too large for conclusions on the composition.
\item With the bremsstrahlung component as derived with the fiducial values for the gas density and the magnetic-field strength, a power-law spectrum in rigidity with index $s=2.25$ provides a reasonable match between the model and the entire gamma-ray SED. A spectral index $s=2.15$ is already at the limit of acceptability, because it overshoots the observed flux above a few hundred GeV, although the hardness of the spectrum can within limits be compensated by choosing a lower maximum energy.
\end{enumerate}
In the absence of a leptonic contribution, the power-law index of the cosmic-ray nuclei should be $s=2.35$, and the first data point at $130~$MeV cannot be reproduced.

\subsection{Secondary electrons}\label{sec-el}
An unavoidable side product in the hadronic scenario is a copious supply of secondary electrons (subsuming electrons and positrons) that result from the production and decay of charged pions. The source rate of electrons is approximately the same as that of gamma rays, but their energy is slightly lower. 
Then, 
GeV-scale gamma-ray emission implies electron production at a Lorentz factor $\gamma_e\approx 10^3$. 
For an elevated production of secondary electrons lasting $10^6\ $yrs or more, M\,82 would be in the calorimetric limit \citep{1989A&A...218...67V}, and the radio flux produced by secondary electrons would reach a constant value. The bolometric nonthermal luminosity would be a direct measure of the cosmic-ray production power. Again using the fiducial parameters (cf. Table~\ref{tab2}), the energy-loss timescale of cosmic-ray nuclei is estimated to be $\tau_\mathrm{loss,pp}\simeq 3\cdot 10^5$~yrs over the energy range between $10$~GeV and $10$~TeV \citep{2007APh....27..429H}. 
Although the starburst activity in M\,82 is not constant \citep{2009ApJ...701.1015K}, the recent spike probably happened about ten million years ago, with strong stellar winds and an elevated supernova rate since then. This is significantly longer than the time needed to reach calorimetry for hadronic and leptonic cosmic rays. Hence, the distribution of secondary electrons should be in the steady state.

The synchrotron flux from the secondary electrons with $E\approx 1$~GeV, or $\gamma_e\simeq 2\cdot 10^3$, that is observable at about $3$~GHz in frequency, or $E_\gamma\approx 10^{-17}\ $TeV, would have to be a factor of approximately 15 weaker than the $\nu F_\nu$ flux of the pion-decay gamma rays at $1$--$2$~GeV. This factor 15 is the product of a factor five for the fraction of energy loss, a factor two for the synchrotron Jacobian, $dE/dE_\gamma$, and a factor $1.5$ for the efficiency ratio of gamma rays and electrons. The observed flux is about at this expected value, indicating that a large fraction of the GeV-scale electrons in M\,82 may be secondary. The following investigates this in more detail to infer whether additional loss processes could be at play. 

To explore the contribution of secondary electrons to the radio spectrum at all frequencies, their production rate, $Q_e$, is calculated with the same model \citep{2020APh...12302490B} used in section~\ref{had-sce} to determine the emissivity of hadronic gamma rays  \citep[see also][]{2008APh....29..282H}. For simplicity, only a spectral index $s=2.25$ and ISM composition is considered, corresponding to the dark-red curve in Figure~\ref{fig:model}. The steady-state electron spectrum, $N=dN/dE$, is well described by the continuity equation
\begin{equation}
 \frac{\partial}{\partial E}\left(b(E)\,N\right)+\frac{N}{T}=Q_e(E)\ .
\label{eq6}
\end{equation}
Here $T$ is the timescale of catastrophic losses that remove particles, for example escape in the wind. For simplicity, it is assumed to be independent of energy. 
The solution to equation~\ref{eq6} is well known \citep{1962SvA.....6..317K},
\begin{equation}
N(E,t)=\int_E^\infty dE'\ \frac{Q_e(E')}{|b(E)|}\,\,\exp\left(-\int_E^{E'} \frac{du}{T\,|b(u)|}\right)\ ,
\label{eq6a}
\end{equation}
where $b(E)$ is the energy loss rate (Equation~\ref{eq7}).

The omnidirectional synchrotron emissivity of a single electron can be written in monochromatic approximation as
\begin{equation}
P_{E_\gamma}= \frac{E^2}{(10^{16}\ \mathrm{GeV}\,\mathrm{s})} \,U_\mathrm{mag}\,\delta\left(E_\gamma -aE^2\right)\ ,
\label{eq8}
\end{equation}
where 
\begin{equation}
a = \left(4\cdot 10^{-16}\ \mathrm{GeV}^{-1}\right)\,\sqrt{U_\mathrm{mag}}\ .
\label{eq9}
\end{equation}
The emission coefficient, $j_{{E_\gamma}}$, for synchrotron radiation then is calculated as
\begin{align}
E_\gamma j_{{E_\gamma}} &= E_\gamma \int dE\ P_{E_\gamma}\,N(E,t) \label{eq10}
\\
 &=\frac{1}{1+\frac{U_\mathrm{rad}}{U_\mathrm{mag}}}
 \frac{E_1^2}{2\left[E_1+E_{c2}\left(1+\frac{E_{c1}}{E_1}\right)\right]}
 \,\mathbf{I} 
 \ ,\nonumber
\end{align}
where
\begin{equation}
\mathbf{I}= \int_{E_1}^\infty dE'\ Q_e(E')\,
\exp\left(-\int_{E_1}^{E'} \frac{du}{T\,|b(u)|}\right)
\end{equation}
and
\begin{equation}
E_1= \sqrt\frac{E_\gamma}{a}.
\end{equation}

\begin{figure}[ht!]
\centering
\includegraphics[width=\columnwidth]{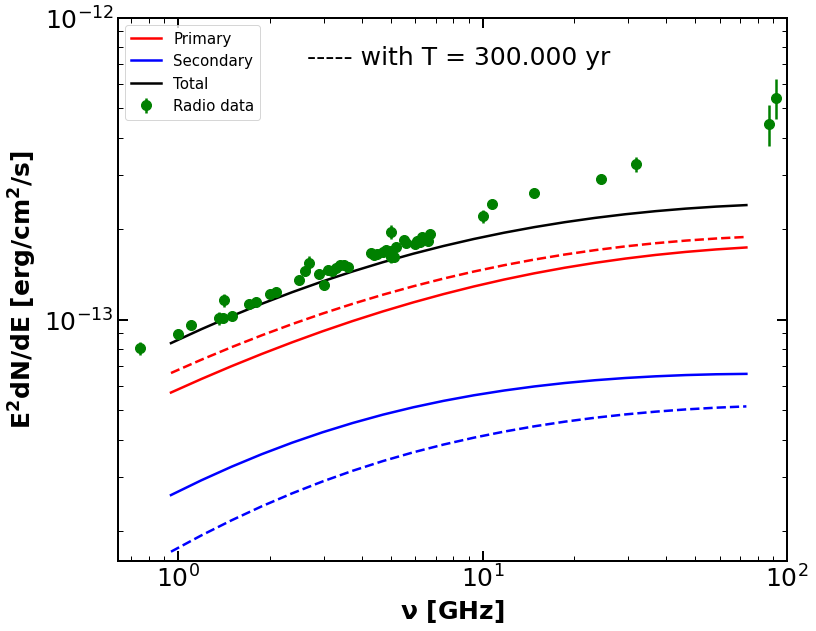}
\caption{Radio SED of M\,82 compared with synchrotron emission spectra from secondary electrons and positrons for the baseline hadronic models (blue lines) with spectral index $s=2.25$ and ISM composition. The solid black line indicates the total synchrotron flux (the same as the blue line in figure~\ref{fig:radio-model-brems}), and the red lines reflect the contribution of primary electrons. Dashed lines show spectra calculated accounting for catastrophic losses with timescale $T=3\cdot 10^5$~yrs. 
\label{fig:radio-model}}
\end{figure}
Figure~\ref{fig:radio-model} compares radio-flux measurements from the literature with the expected synchrotron spectra of secondary electrons and positrons alone (Equation~\ref{eq10}). The source rate of secondary electrons, $Q_e$ in equation~\ref{eq6}, corresponds to the gamma-ray emission model with cosmic-ray spectral index $s=2.25$ and ISM composition, that in Figure~\ref{fig:model} was shown to very well match the observed GeV/TeV spectrum.

The synchrotron flux from secondary electrons in the calorimetric limit is about a third of the observed flux at a few GHz, approximately scaling with the inverse of the scaling factor for the bremsstrahlung flux (cf. equation~\ref{eq_scalebr}). As previously noted in section~\ref{had-sce}, this factor cannot significantly exceed unity, and in consequence the scaling factor for the radio flux from secondary electrons may not significantly fall below unity.

This leaves two thirds or less of the radio flux to be provided by primary electrons. There is little spectral difference between the synchrotron emission from primary and secondary electrons, largely because their source spectra are found to be similar above a GeV. 
The inhomogeneity in the starburst core makes estimating a mean magnetic field and mean gas density difficult, and it may be that the magnetic field is substantially weaker than the fiducial value, or the gas density much higher than that given in table~\ref{tab2}, reducing the synchrotron emission from secondary electrons. However, this would imply more GeV-scale electrons in the system, and hence a larger bremsstrahlung flux that would exceed the observed gamma-ray flux at $100~$MeV -- $500~$MeV. 

An alternative option is another loss process on a similar time scale that is not accounted for here. Figure~\ref{fig:radio-model} also contains the expected radio spectrum from secondary electrons in the calorimetric limit with an additional catastrophic loss on the timescale $T=3\cdot 10^5$~yrs, for which the contribution of primary electrons to the GHz-scale synchrotron flux could rise to about 75\%. Such a loss process may be advection in the starburst wind, in which localized driving and radiative cooling lead to a multi-phase structure composed of filaments of dense gas embedded in a very dilute gas that flow at high speed \citep[e.g.][]{10.1093/mnras/stt126}. Advective transport out of the starburst core over a distance of $300$~pc in a wind moving at $1000~ \mathrm{km/s}$ would take $\tau_\mathrm{adv}\simeq 3\cdot 10^5$~yrs, sufficient to significantly reduce the radio emission from secondary electrons.

\subsection{The cosmic-ray spectrum}\label{sed:discussion}
If limited by available time and not by any loss process, the spectrum of cosmic rays should reflect the spectrum with which the particles are produced. Otherwise, the losses may modify the spectrum. Again using the aforementioned fiducial parameters (cf. Table~\ref{tab2}), the energy-loss timescale of cosmic-ray nuclei is estimated to be $\tau_\mathrm{loss,pp}\simeq 3\cdot 10^5$~yrs with little variation over the energy range between $10$~GeV and $10$~TeV \citep{2007APh....27..429H}. For cosmic-ray electrons, $\tau_\mathrm{loss,e}\simeq 10^5$~yrs at $1$~GeV, corresponding to synchrotron radiation at a few GHz, with a transition to a decreasing lifetime above that energy. In the previous section, it was shown that additional loss processes, possibly advective escape, may be at play with a lifetime similar to that of the energy losses. 

A complication for advective escape lies in the diffusive transport between the dense filaments, where most of the radiation is produced, and the dilute, fast-flowing gas that provides advective transport. To our knowledge, there is no assessment of cosmic-ray spectra in a complex environment like that in M\,82. There are analytical 1-D estimates of cosmic-ray transport in an accelerating wind, which suggest that even in the steady state for losses by advection the cosmic-ray spectrum would be the production spectrum steepened by one-half of the energy dependence of the diffusion coefficient
\citep{1982MNRAS.201.1041L}, and so the observed spectral index must be close to the mean production spectral index. For cosmic-ray ions in M\,82 we found a spectral index $s=2.25$. The radio spectrum, for comparison, is compatible with a source spectral index $s\simeq 2.25$ for electrons in the energy range $1-10$~GeV. 

The production spectral index around $s\simeq 2.25$ deduced for the cosmic rays in M\,82 is somewhat softer than that provided by diffusive shock acceleration. Simulations suggesting a spectral softening on account of different downstream advection speeds of gas and magnetic turbulence have been reported \citep{2020ApJ...905....2C}, but have not been independently confirmed to date. Energy loss by driving of turbulence has been shown to have very little, if any, effect on the spectrum \citep{2021ApJ...921..121P}.

The particle spectrum produced at a certain time and the time-integral of the production spectrum are two different quantities. Relevant for the analysis offered in this paper is the latter, and diffusive shock acceleration provides the former. Studies of supernova remnants indicate that with time the interplay of turbulence driving in the upstream region and particle acceleration at the shock becomes less efficient, primarily resulting in a reduction of the maximum particle energy that can be reached \citep{2019MNRAS.490.4317C,2020A&A...634A..59B}. The time-integral of the production spectrum would be slightly softer than what we deduced for M\,82 above a few GeV, with an effective spectral index of s$\approx 2.4$. 

The supernova remnants and the wind bubbles of massive stars in M\,82 are likely overlapping, leading to a complex and turbulent structure composed of hot, dilute plasma and cool, dense gas organized in shells \citep{2022MNRAS.517.2818B}. The situation resembles a supernova remnant in the wind bubble of a Wolf-Rayet progenitor \citep[e.g.][]{2020MNRAS.493.3548M} and interacting supernova remnants like IC443 \citep{2021A&A...649A..14U}. If a supernova shock passes through plasma that has been shocked before by another supernova or a wind termination shock, then the temperature is very high and the sonic Mach number low, resulting in a soft particle spectrum with low cosmic-ray density \citep[e.g.][]{2022A&A...661A.128D}. If the supernova shock hits a filament of dense gas or the outer shell of a collective wind bubble, it splits into a reflected shock, that provides little acceleration to high energies, as well as a transmitted shock that will propagate through the dense gas. Many particles may be accelerated by the transmitted shock, but its speed is low, and so is the maximum energy to which it can accelerate. 

\section{Summary}
The VERITAS collaboration performed a long-term study of the starburst galaxy M\,82 at GeV--TeV energies using more than 10 years of data. This study primarily led to a better measurement of the object's VHE flux and its photon spectrum. Combining  
the improved VHE spectrum with multi-wavelength data at lower energies, enables robust spectral modelling. The 
results of the spectral modelling are summarized below.

\begin{itemize}
\item Both leptonic and hadronic scenarios are considered to explain the observed SED, and the hadronic model is clearly preferred. The gamma-ray SED is compatible with cosmic-ray nuclei following a momentum spectrum with index $s\simeq 2.25$. Models with a heavy composition of both cosmic rays and gas were tested and found to be a reasonable match of the data as well, suggesting that gamma-ray-based statements on a possible elemental enrichment of the material in the starburst core are not possible at this time.
\item The observed gamma-ray flux at $100~$MeV to $500~$MeV is likely dominated by bremsstrahlung, which implies an electron-to-ion ratio of GeV-scale electrons in M\,82 that is much larger than that in, e.g., the Milky Way Galaxy, on account of the near-equality of the energy-loss timescales for pion production and bremsstrahlung.
\item Primary electrons and secondary electrons associated with the protons can explain the observed radio spectrum while not overproducing the diffuse X-ray emission, which is only partly nonthermal. Additional loss processes besides the radiative energy losses of electrons may be at play; otherwise the GHz-scale radio flux from the secondary electrons would be very close to that from primary electrons. Advective escape in the galactic wind could be that additional loss process. A significant reduction of the radio flux from the secondary electrons by, e.g., a higher gas density than assumed here is not possible, because the implied bremsstrahlung flux at a few hundred MHz would exceed the observed level.
\item The lifetime of cosmic rays in M\,82 is much shorter than the duration of the starburst, suggesting calorimetric behavior. If a significant loss channel for electrons is non-radiative, for example by escape, then the radiation output is only partially calorimetric, and probably likewise for cosmic-ray nuclei.
\item The soft spectral index of $s\simeq 2.25$ for the cosmic-ray source spectrum that the spectral modelling seems to prefer is in line with results for individual supernova remnants. Among the possible reasons for spectra softer than the test-particle limit of diffusive shock acceleration are inefficient turbulence driving upstream of aged shocks, the high temperature of previously shocked upstream plasma, or the small speed of shocks transmitted into filaments of dense gas in the starburst core.
\end{itemize}

\section{Acknowledgments}
This research is supported by grants from the U.S. Department of Energy Office of Science, the U.S. National Science Foundation and the Smithsonian Institution, by NSERC in Canada, and by the Helmholtz Association in Germany. This research used resources provided by the Open Science Grid, which is supported by the National Science Foundation and the U.S. Department of Energy’s Office of Science, and resources of the National Energy Research Scientific Computing Center (NERSC), a U.S. Department of Energy Office of Science User Facility operated under Contract No. DE-AC0205CH11231. We acknowledge the excellent work of the technical support staff at the Fred Lawrence Whipple Observatory and at the collaborating institutions in the construction and operation of the instrument.


\bibliography{masterbibtex}{}
\bibliographystyle{aasjournal}



\end{document}